\documentclass[%
 aip,
 amsmath,amssymb,
 reprint,%
]{revtex4-1}

\usepackage{graphicx}
\usepackage{dcolumn}
\usepackage{bm}
\usepackage{xcolor}
\usepackage[utf8]{inputenc}
\usepackage[T1]{fontenc}
\usepackage{mathptmx}
\usepackage{etoolbox}
\usepackage{gensymb}
\usepackage{hyperref}
\usepackage{xr-hyper}

\begin{document}

\title{Interfacial Control of Orbital Occupancy and Spin State in LaCoO$_3$} 

\author{Ellen M. Kiens}
\email{e.m.kiens@utwente.nl}
\affiliation{MESA+ Institute for Nanotechnology, Faculty of Science and Technology, University of Twente, 7500 AE Enschede, The Netherlands}

\author{Nicolas Gauquelin}
\affiliation{Electron Microscopy for Materials Science, University of Antwerp, Campus Groenenborger,
Groenenborgerlaan 171, 2020 Antwerpen, Belgium}

\author{Arno Annys}
\affiliation{Electron Microscopy for Materials Science, University of Antwerp, Campus Groenenborger,
Groenenborgerlaan 171, 2020 Antwerpen, Belgium}

\author{Emma van der Minne}
\affiliation{MESA+ Institute for Nanotechnology, Faculty of Science and Technology, University of Twente, 7500 AE Enschede, The Netherlands}

\author{Iris C.G. van den Bosch}
\affiliation{MESA+ Institute for Nanotechnology, Faculty of Science and Technology, University of Twente, 7500 AE Enschede, The Netherlands}

\author{Matthijs A. van Spronsen}
\affiliation{Diamond Light Source, Harwell Science and Innovation Campus, Didcot, Oxfordshire OX11 0DE, United Kingdom}

\author{Zezhong Zhang}
\affiliation{Electron Microscopy for Materials Science, University of Antwerp, Campus Groenenborger,
Groenenborgerlaan 171, 2020 Antwerpen, Belgium}

\author{Annick de Backer}
\affiliation{Electron Microscopy for Materials Science, University of Antwerp, Campus Groenenborger,
Groenenborgerlaan 171, 2020 Antwerpen, Belgium}

\author{Sandra van Aert}
\affiliation{Electron Microscopy for Materials Science, University of Antwerp, Campus Groenenborger,
Groenenborgerlaan 171, 2020 Antwerpen, Belgium}

\author{Jo Verbeeck}
\affiliation{Electron Microscopy for Materials Science, University of Antwerp, Campus Groenenborger,
Groenenborgerlaan 171, 2020 Antwerpen, Belgium}

\author{Gertjan Koster}
\affiliation{MESA+ Institute for Nanotechnology, Faculty of Science and Technology, University of Twente, 7500 AE Enschede, The Netherlands}

\author{Bastian Mei}
\affiliation{MESA+ Institute for Nanotechnology, Faculty of Science and Technology, University of Twente, 7500 AE Enschede, The Netherlands}
\affiliation{Laboratory of Industrial Chemistry, Ruhr University Bochum, 44780 Bochum, Germany}

\author{Frank M.F. de Groot}
\affiliation{Materials Chemistry and Catalysis, Debye Institute for Nanomaterials Science, Utrecht University, Universiteitsweg 99, 3584 CG Utrecht, The Netherlands}

\author{Christoph Baeumer}
\email{c.baeumer@utwente.nl}
\affiliation{MESA+ Institute for Nanotechnology, Faculty of Science and Technology, University of Twente, 7500 AE Enschede, The Netherlands}
\affiliation{Peter Gruenberg Institute 7, Forschungszentrum Juelich GmbH, 52428 Juelich, Germany}

\date{\today}

\begin{abstract}
Transition metal oxides exhibit a wide range of tunable electronic properties arising from the complex interplay of charge, spin, and lattice degrees of freedom, governed by their \textit{d} orbital configurations, making them particularly interesting for oxide electronics and (electro)catalysis. Perovskite oxide heterointerfaces offer a promising route to engineer these orbital states. In this work, we tune the Co 3\textit{d} orbital occupancy in LaCoO$_3$ from a partial $d^7$ to a partial $d^5$ state through interfacial engineering with LaTiO$_3$, LaMnO$_3$, LaAlO$_3$ and LaNiO$_3$. Using X-ray absorption spectroscopy combined with charge transfer multiplet calculations, we identify differences in the Co valence and spin state for the series of oxide heterostructures. LaTiO$_3$ and LaMnO$_3$ interfaces result in interfacial charge transfer towards LaCoO$_3$, resulting in a partial $d^7$ orbital occupancy, while a LaNiO$_3$ interface results in a partial Co $d^5$ occupancy. Strikingly, a LaAlO$_3$ spacer layer between LaNiO$_3$ and LaCoO$_3$ results in a Co $d^6$ low spin state. These results indicate that the Co spin state, like the valence state, is governed by the interfacial environment. High-resolution scanning transmission electron microscopy imaging reveals a clear connection between strain and spin configuration, emphasizing the importance of structural control at oxide interfaces. Overall, this work demonstrates that interfacial engineering simultaneously governs orbital occupancy and spin state in correlated oxides, advancing spin-engineering strategies in correlated oxides and offering new insights for the rational design of functional oxide heterostructures.
\end{abstract}

\maketitle 

\section{\label{sec:introduction}Introduction}
\begin{figure*}[t]
\includegraphics[width=\textwidth]{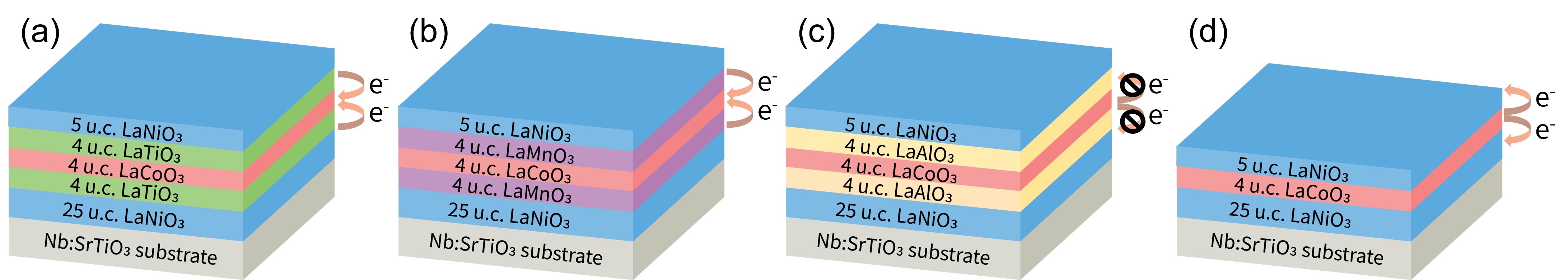}
\caption{\label{fig:LCOsamples} Family of samples to study the four La\textit{B}O$_3$-LaCoO$_3$ interfaces, where \textit{B} is (a) Ti, (b) Mn, (c) Al, (d) Ni.}
\end{figure*}

Transition metal oxides are of great technological importance due to their wide range of tunable electronic properties. The ordering and occupation of the outer \textit{d} orbitals in these materials results in their structural, electronic, magnetic, optical and thermal properties. Moreover, the strongly correlated \textit{d} electrons in these materials entangle charge, spin and structural degrees of freedom, giving rise to a variety of exotic phenomena such as magnetic ordering, metal-to-insulator transitions, multiferroics and superconductivity \cite{Tokura_orbitalphysics,Goodenough_perspectiveTMOs}. Engineering of the orbital configuration in these highly correlated materials can lead to new emerging physical properties.\\ Perovskite oxide heterointerfaces offer a platform to tune orbital configurations and thereby explore new electronic phenomena\cite{mannhartschlom_science,Hwang2012_natmater}. For example, by structural engineering, exploiting strain and oxygen octahedral coupling across the interface, electronic and magnetic properties can be tuned\cite{HuijbenNatureMat2016,Liu_SROLNO_magnetism,Li_LMOSTO_magnetism,Ji_LCOLSMO,Guan_LMOLCO_magnetism_octahedralcoupling,Hoffman_chargetransfer_LNOLMO,Cao_LNOLTO_CT_Mottgroundstate,Disa_LTOLNO,Chen_CTreview,LCOSTOsuperlattices}. Additionally, when the materials on both sides of the interface have different electron chemical potential, charge transfer can take place. For epitaxial isopolar heterointerfaces, the oxygen octahedra form a continuous backbone throughout the heterostructure. This should cause the O $2p$ levels of both materials across the interface to align. To align the chemical potential of both materials, charge transfer across the interface is expected \cite{ZhongHansmann}. This picture differs from the extensively studied polar interface, where a polar overlayer and/or the creation of point defects give rise to different electronic states at the interface \cite{Huijben_LAOSTO}.\\
Interfacial charge transfer across isopolar interfaces presents an alternative method to control the \textit{d} orbital occupation to chemical doping. For example, divalent Co, Fe and Ni have been observed at the interface between LaCoO$_3$, LaFeO$_3$ and LaNiO$_3$ with LaTiO$_3$\cite{JaapLTOLCO,KleibeukerLTOLFO,Cao_LNOLTO_CT_Mottgroundstate}. Following the band alignment picture, the oxidation state of Co, Fe and Ni at these interfaces were attributed to interfacial electron transfer from Ti to the heavier transition metals. However, the experimental study of LaTiO$_3$ interfaces is complex due to the propensity of LaTiO$_3$ to overoxidize \cite{ScheidererLTO,OhtomoLTO}.\\
Tuning the \textit{d} orbital occupation of Co in LaCoO$_3$ is of particular interest for two reasons. First, the intriguing spin state of LaCoO$_3$ and its related ferromagnetic order has been extensively discussed in literature \cite{Korotin_intermediatespin,Bhide_LCOspin,Kyomen_LCOspin,Naiman_LCOspin,Raccah_LCOspin,Podlesnyak_LCOspin,HaverkortLaCoO3,RupanLaCoO3,Sterbinsky_LCOexafs,Hsu_LCOepitaxialcalculations,Gupta_LCOstraindirvenmagnetism}. The ground state of Co in LaCoO$_3$ is $d^6$ and its spin state is dictated by the competition between the crystal field splitting energy between the 3\textit{d} orbitals into e$_g$ and t$_{2g}$ levels and the exchange energy. Bulk LaCoO$_3$ exhibits a low spin state at low temperature (20 K) and an inhomogeneous mixed spin state at room temperature \cite{HaverkortLaCoO3}. However, for thin films the spin state differs from the bulk single crystal due to epitaxial strain \cite{RupanLaCoO3,Sterbinsky_LCOexafs,Hsu_LCOepitaxialcalculations,Gupta_LCOstraindirvenmagnetism} and reduced dimensionality \cite{LAOLCOsuperlattice}.\\
Second, Co based transition metal oxides are promising materials for (electro)catalysis. The \textit{d} orbital occupation of perovskites has been identified as a descriptor of their performance in catalyzing the oxygen evolution reaction, which is the limiting half-reaction in electrochemical water splitting \cite{Suntivich_egdescriptor}. LaCoO$_3$ is a promising electrocatalyst, for which control over the $d$ orbital occupation might allow for tuning of its (electro)chemical reactivity. Lowering its \textit{d} orbital occupation to a (partial) $d^5$ ground state could result in a highly active electrocatalyst\cite{nickelatesfeaturearticle,HeymannLSCOLCO}.
Although interfacial charge transfer at LaCoO$_3$-based interfaces has been reported previously \cite{JaapLTOLCO}, a systematic investigation of charge transfer in both directions, as well as the connection between orbital occupancy and structural effects across different interfaces remains absent.\\ 
In this work, we aim to tune the $d$ orbital occupation of Co in LaCoO$_3$ by interfacial engineering from a partial $d^5$ to a partial $d^7$ state by fabrication of the interfaces shown in \autoref{fig:LCOsamples}. Following the band alignment picture proposed by Zhong and Hansmann \cite{ZhongHansmann}, a partial $d^7$ occupation is expected at the interface with LaTiO$_3$ and LaMnO$_3$ while a partial $d^5$ occupation is expected at the interface with LaNiO$_3$. Furthermore, we investigate blocking of charge transfer for the latter using a LaAlO$_3$ spacer layer. We studied the Co electronic state by X-ray absorption spectroscopy (XAS). The relative contributions of different valence and spin states are attributed with the help of charge transfer multiplet (CTM) calculations. Next, we examined the intermixing effects and strain in the LaCoO$_3$ layer for the different interfaces with scanning transmission electron microscopy (STEM). We find that Co valence and spin state are influenced by interfacial engineering and provide evidence that structural effects and strain are governing the $d$ orbital occupation at these oxide interfaces. 

\section{\label{sec:methods}Methods}
\subsection{Experimental}
Fully strained epitaxial thin films were grown with unit cell thickness control using pulsed laser deposition (PLD) with in-situ reflective high-energy electron diffraction (RHEED). The depositions were performed in a vacuum chamber with a base pressure of 5x10$^{-8}$ mbar, equipped with in-situ RHEED (Staib instruments) and a KrF excimer laser with a wavelength of 248 nm (Coherent). The deposition parameters (Table \ref{tab:deppar}) were varied slightly per material to facilitate layer-by-layer growth and thereby unit cell thickness control. After growth, the multilayers were cooled down in deposition pressure with 25 $\degree$C/min. Each epitaxial multilayer was grown on TiO$_2$ terminated \cite{Koster_STOtreatment} Nb:SrTiO$_3$ 0.5 wt\% substrates in the (001) orientation  (CrysTec GmbH, Germany). A 25 u.c. LaNiO$_3$ was grown on top of the substrate to separate the interface of interest from the polar interface with SrTiO$_3$, and to ensure similar rotation and/or tilting of oxygen octahedra between samples\cite{FowlieLNOrotations}. A relatively low growth temperature (450 $\degree$C) is chosen for the LaNiO$_3$ to obtain B-site termination of this layer \cite{Baeumer_LNOnatmater}. The thickness of the LaCoO$_3$ layer was chosen to be 4 u.c., as interfacial charge transfer is expected to extend 2-3 u.c. from the interface as previously reported by Araizi-Kanoutas \textit{et al.}\cite{JaapLTOLCO}. The thickness of the La\textit{B}O$_{_3}$ ($B$ = Ti, Mn, Ni) layers below and on top of the LaCoO$_3$ were chosen to be just above this limit and therefore 4 u.c. as well. A 5 u.c. LaNiO$_3$ capping layer on all samples prevented beam damage during XAS measurements and air exposure effects.\\
X-ray diffraction was used to obtain structural information on the macroscopic scale. Diffractograms were obtained using a Bruker D8 Discover with a rotating anode source, Montel optics and a 2D Eiger2 R 500K detector. A two-bounce channel-cut germanium monochromator was used to obtain Cu-K$\alpha$ radiation and a 1 mm collimator was used for a well-defined beam size.\\
The morphology of the multilayers was examined by atomic force microscopy (AFM) using a Veeco Dimension Icon AFM in tapping mode in air with a Tespa-V2 cantilever (Bruker, Netherlands) and a silicon tip with a nominal tip radius of 20 nm.\\
The $d$ orbital occupation of the 3$d$ transition metals in the multilayers was studied by X-ray absorption spectroscopy (XAS) of the transition metal $L$ edges. These experiments were performed at the Versatile Soft X-ray (VerSoX) beamline B07-b at Diamond Light Source \cite{B07beamline}. Spectra were collected in total electron yield at room temperature in vacuum (10$^{-8}$ mbar). The incident angle of the photons was perpendicular to the sample surface. We took at least three spectra at different positions per sample. Furthermore, an hour of consecutive scans on the LaNiO$_3$-LaCoO$_3$ multilayer shows no change in spectral shape upon X-ray radiation for Co. The spectra presented in this work are single scans. As no spatial or time-dependent changes were observed, these scans are assumed to be representative of the whole sample. For all spectra, a linear background was subtracted well before the edge after which the spectra were normalized to the edge jump, thereby accounting for the number of holes in the 3$d$ shell.\\
Focused ion beam lamellas for Scanning Transmission Electron Microscopy (STEM) were prepared in an Helios Nanolab 650 in a Kamrath and Weiss vacuum transfer box and transferred directly to a Gatan TEM vacuum transfer holder as described elsewhere \cite{FIBref_reclaiming,FIB_hybridperovskite,FIBref_berryphase,JaapLTOLCO}.\\
STEM using High Angle Annular Dark Field (HAADF) imaging was performed using a FEI Titan 80-300 microscope operated at 300 kV. Electron Energy Loss Spectroscopy (STEM-EELS) measurements were performed using a monochromatic beam with a 200 meV energy resolution and an acceleration voltage of 120 kV on a FEI Titan 80-300 microscope. The Ti $L$, Co $L$, Mn $L$, Ni $L$ edge, O $K$ and La $M$ edges were acquired in several measurements. The acquisition parameters were 0.2 s/pixel, 0.12 Å/pixel and 0.05 eV/pixel in the dual EELS mode. Collection angles for HAADF imaging and EELS were 70-160 mrad and 47 mrad, respectively.\\
EELS analysis was performed using the pyEELSMODEL software \cite{pyEELS}. The Co $L$ edge, Ti $L$ edge, Mn $L$ edge and O $K$ edge traces were computed using the background removal routine. The Ni $L$ edge traces were calculated by model-based fitting because of the strong overlap with the La $M$ edge. The model consisted of a constrained background component \cite{EELSbackground}, a La $M$ reference edge extracted from a Ni-free region and a model Ni $L$ edge using the theoretical cross-section from Z. Zhang \textit{et al.} \cite{EELScrossection} combined with a generic fine structure component.\\
For Energy-dispersive X-ray spectroscopy (EDX) elemental mapping, multi-frame EDX data was acquired, using the simultaneously acquired HAADF signal. All frames were aligned using the SuperAlign routine \cite{EDX_superalign}.\\
Strain maps were generated using the StatSTEM software \cite{StatSTEM}. In this approach, the images are modeled as a superposition of Gaussian functions centered on the atomic columns, allowing the positions of these columns to be determined with high accuracy and precision. The displacement field was obtained by comparing the measured atomic column positions with those of an ideal, undistorted reference lattice. The reference lattice is defined by the SrTiO$_3$ substrate at the bottom of the structure. Specifically, the projected lattice vectors along the \textbf{a} and \textbf{b} directions are determined from the measured atomic column positions of SrTiO$_3$. These vectors are then used to construct a reference lattice, which serves as the basis for measuring strain throughout the entire layered structure. These displacement vectors were then projected onto the local lattice basis defined by the crystallographic directions \textbf{a} and \textbf{b}. For each analysis region, the deformation gradient tensor ($F$) was calculated from the reference and deformed displacement fields. The Cauchy strain tensor ($\epsilon$) was subsequently computed according to $\epsilon = \frac{1}{2}(F + F^T)-I$, where $F^T$ is the transpose of the deformation gradient tensor $F$ and $I$ is the identity tensor. This method yields the strain components with atomic resolution.

\subsection{Charge transfer multiplet calculations} 
Charge transfer multiplet calculations based on the Cowan-Butler-Thole code \cite{cowantheory,Thole,butlerpoint} were performed to attribute the valence contributions of each spectrum. The spectra were calculated from the sum of all possible excitations of an electron from the occupied 2$p$ to unoccupied 3$d$ states ($2p^6$$3d^n$ $\rightarrow$ $2p^5$$3d^{n+1}$). The ground state ($2p^6$$3d^n$) is affected by the crystal field, 3$d$ spin-orbit coupling, and 3$d$-3$d$ electron interaction (multiplet coupling). The final state ($2p^5$$3d^{n+1}$) is affected by the crystal field and 2$p$ and 3$d$ spin-orbit coupling, 2$p$-3$d$ and 3$d$-3$d$ electron interaction (multiplet coupling). We simulated Co-O covalency using ligand-to-metal charge transfer, which transitions can be described as $2p^6$$3d^m$\underline{$L$} $\rightarrow$ $2p^5$$3d^{m+1}$\underline{$L$}, where $m = n+1$ and \underline{$L$} denotes a hole in the oxygen ligand. The atomic Slater–Condon parameters ($F_{dd}$, $F_{pd}$, and $G_{pd}$) were scaled to 80\% of the Hartree–Fock values. The calculations were performed with the CTM4XAS interface \cite{CTM4XAS}. The semi-empirical calculation parameters are listed in supplementary section \ref{sec:CTMcalc}. To estimate the valence state contributions to the experimental Co $L$ edges, linear combination fitting of simulated spectra was performed after subtraction of the step edge. The step edge is typically located a few eV above the white line \cite{VanderLaan_stepedges}. Hence, the step edge was subtracted at 3.5 eV above the white line for Co$^{2+}$, 1.5 eV for Co$^{3+}$,  and 1.5 for Co$^{4+}$ with a ratio of 2:1 for the $L_3$ and $L_2$ edge, respectively. The simulated spectra were normalized to the number of holes in the 3$d$ orbitals before linear combination fitting.

\section{\label{sec:results}Results and discussion} 
\begin{figure*}
\includegraphics[width=\textwidth]{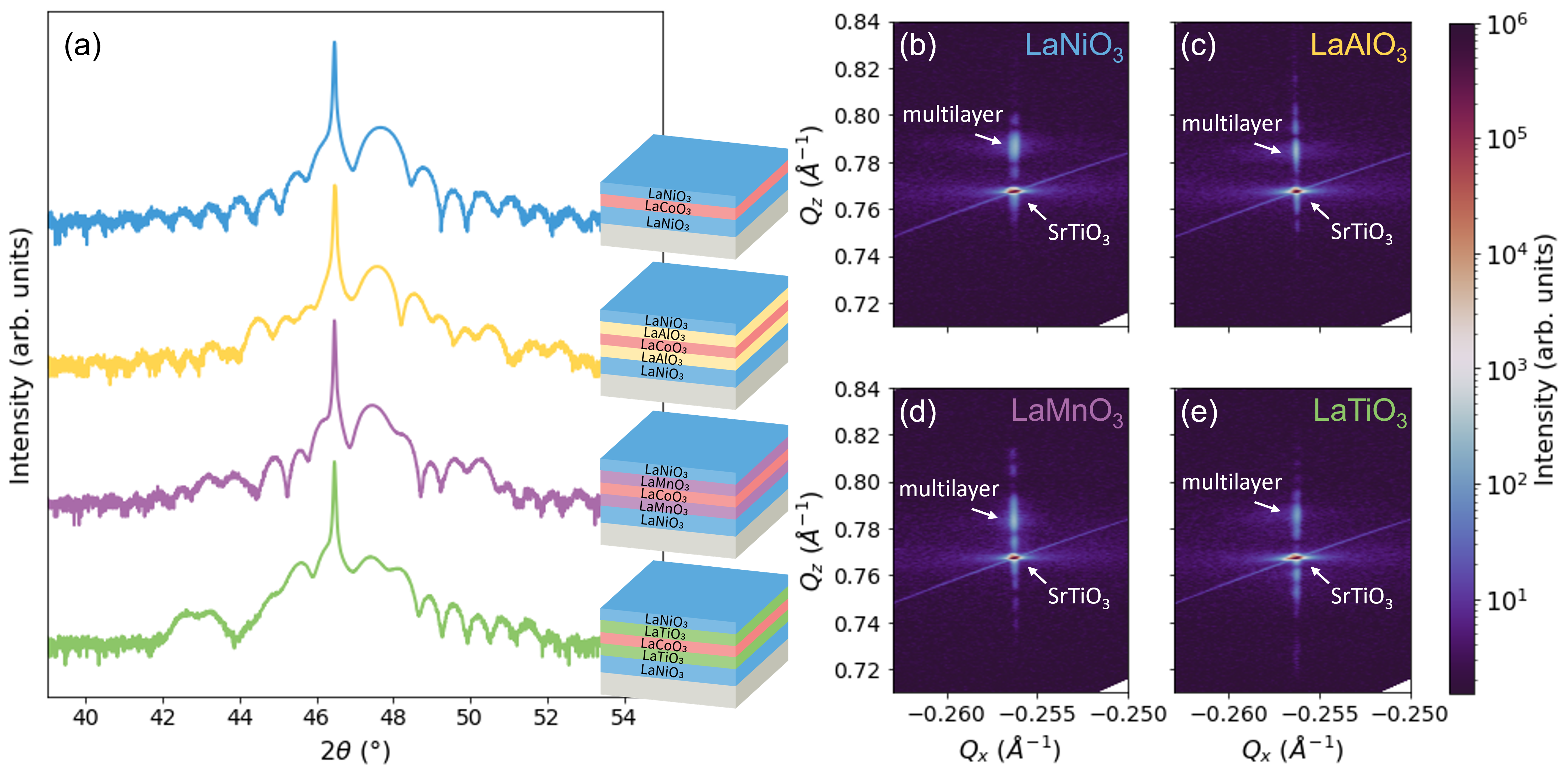}
\caption{\label{fig:XRDresults} X-ray diffraction measurements of the LaCoO$_3$ multilayers: (a) 2$\theta$-$\omega$ scans. (b) Reciprocal space maps of the 103 reflections of the multilayers containing the (b) LaNiO$_3$-LaCoO$_3$ interface, (c) LaAlO$_3$-LaCoO$_3$ interface, (d) LaMnO$_3$-LaCoO$_3$ interface, (e) LaTiO$_3$-LaCoO$_3$ interface.}
\end{figure*}

To assess the structural quality of the multilayers, we applied a comprehensive set of characterization methods. X-ray diffraction measurements (\autoref{fig:XRDresults}) indicate successful growth of highly crystalline epitaxial multilayers. The 2$\theta$-$\omega$ scans (\autoref{fig:XRDresults}a) show a complex oscillatory pattern of Laue fringes around the film peak due to interference effects of the layers with slightly different out-of-plane lattice parameters. The presence of the fringes points towards highly crystalline layers with sharp interfaces at the macroscopic scale \cite{OpticalprinciplesXRD}, consistent with the RHEED patterns (Figure \ref{fig:rheedSI}). Reciprocal space maps around the 103 reflections (\autoref{fig:XRDresults}b) show that all films are commensurately strained to the substrate. In addition, AFM measurements (Figure \ref{fig:AFMresults}) show characteristic vicinal step-terraces, also observed for the substrate, with some unit cell height variations. 

\begin{figure}
\includegraphics[width=0.5\textwidth]{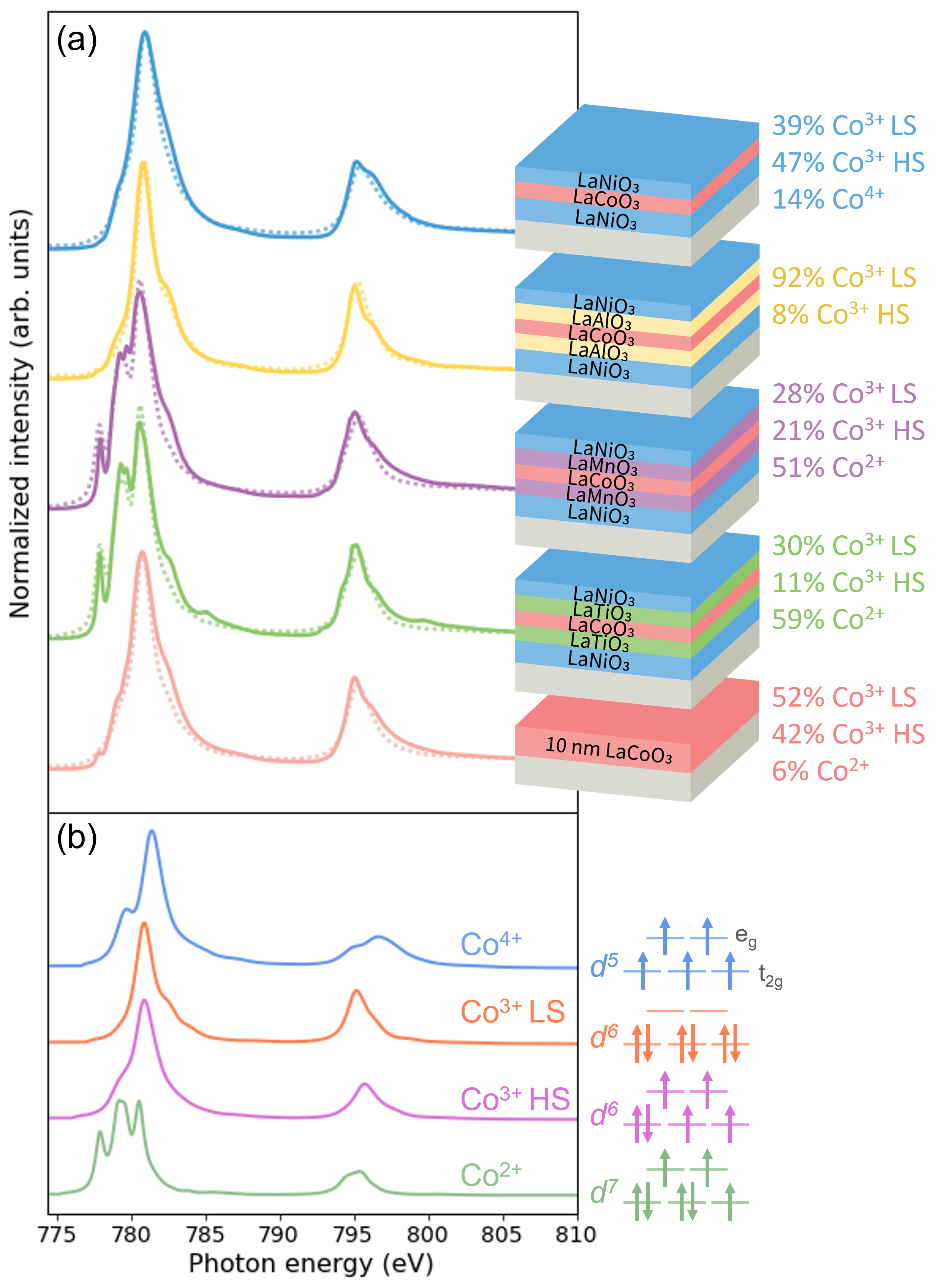}
\caption{\label{fig:XAS_Co} (a) Co L$_{2,3}$ edges for different LaCoO$_3$ and a 10 nm single LaCoO$_3$ film (solid lines). Dashed lines represent a linear combination fit of the simulated spectra. (b) Simulated Co L$_{2,3}$ edges for different valence and spin states.}
\end{figure}

The Co oxidation state is examined by X-ray absorption spectroscopy. The transition metal \textit{L} edges correspond to transitions of electrons from 2$p$ orbitals into empty states. They are dominated by excitations to the empty 3$d$ orbitals ($2p^6$$3d^n$ $\rightarrow$ $2p^5$$3d^{n+1}$), which makes the spectra sensitive to the $d$ orbital occupation of the initial state \cite{DeGroot_XASreview}. \autoref{fig:XAS_Co}a shows the Co $L$ edges of the multilayers and a single 10 nm LaCoO$_3$ layer.  To attribute the contributions of each valence state to the spectra, we simulated the spectral shape of Co$^{2+}$, Co$^{3+}$ and Co$^{4+}$ using charge transfer multiplet calculations (\autoref{fig:XAS_Co}b). For Co$^{2+}$, we only consider a HS state as the LS state is induced only by strong Jahn-Teller distortions \cite{Hernandez_Co2LS}. As Co in LaCoO$_3$ is reported to contain a mixture of Co$^{3+}$ LS and HS \cite{HaverkortLaCoO3}, both spin states are considered. For Co$^{4+}$ we consider only a HS contribution as the experimental spectrum of SrCoO$_3$ looks most similar to HS Co$^{4+}$ \cite{Potze_SrCoO3}. In the HS configuration of $d^5$, each orbital would be occupied by one electron in each $d$ orbital, which is considered to be a likely configuration. A linear combination fitting of these simulated spectra helps to identify differences between samples. However, due to differences between theory and experiment, the concentrations extracted from linear combination fitting serve only as approximate indicators and should not be interpreted as exact values of the Co oxidation states within each sample.\\
The spectrum obtained for a single 10 nm LaCoO$_3$ film, shown in red in \autoref{fig:XAS_Co}, shows features of multiple oxidation states. Based on the stoichiometry, a Co$^{3+}$ oxidation state ($d^6$) is expected for the single layer. While the position of the white line and the shoulder at $\sim$782 eV are characteristic for a Co$^{3+}$ oxidation state, an additional feature is visible at $\sim$778 eV. This feature indicates a small concentration of Co$^{2+}$, often observed for LaCoO$_3$ surfaces\cite{MerzLCOLCCOLSCO,PintaLCO,JaapLTOLCO,AbbateLCO}. This small amount of Co$^{2+}$ is likely due to defects in the surface region. Haverkort \textit{et al.} found that LaCoO$_3$ is an inhomogeneous mixed-spin state system that can be simulated by the sum of Co$^{3+}$ LS and HS states. By fitting a linear combination of the simulated spectra for Co$^{3+}$ LS and HS and Co$^{2+}$ (the dotted line in \autoref{fig:XAS_Co}a), we find a slightly higher contribution of HS compared to LS and approximately 6\% Co$^{2+}$. This is a larger HS population than reported for LaCoO$_3$ single crystals \cite{HaverkortLaCoO3}, but similar to other LaCoO$_3$ epitaxial thin films \cite{MerzLCOLCCOLSCO,PintaLCO}. This higher HS population is expected due to the epitaxial strain in the thin film \cite{RupanLaCoO3}.\\ 
When creating a LaNiO$_3$-LaCoO$_3$ interface, the white line of the $L_3$ shifts $\sim$0.2 eV to higher photon energies as shown in blue in \autoref{fig:XAS_Co}. Furthermore, the shoulder at $\sim$782 eV is less pronounced. Based on the predictive spectral shapes using CTM calculations (\autoref{fig:XAS_Co}b), the LaNiO$_3$-LaCoO$_3$ interface could either be the result of a different Co$^{3+}$ LS/HS population, or contain some Co$^{4+}$ contribution. The shift of the white line is indicative of a higher oxidation state. The difference in the $L_3$ white line position is $\sim$0.3 eV for LaCoO$_3$ (Co$^{3+}$) and SrCoO$_3$ (Co$^{4+}$) \cite{Harvey_BSCFO,Li_LCOSCOref}. Therefore, the small shift for the LaNiO$_3$-LaCoO$_3$ interface points towards some Co$^{4+}$ population. A linear combination fit with and without the Co$^{4+}$ component shown in Figure \ref{fig:XASCofits}b and c, respectively. The linear combination fit without the Co$^{4+}$ component shows a mismatch in the white line position as well as a discrepancy in spectral shape. Including a small amount of Co$^{4+}$ ($\sim$ 14\%) results in a better match with the experimental data. This points towards a partial Co $d^5$ orbital occupation in proximity of the LaNiO$_3$-LaCoO$_3$ interface. If this partial $d^5$ occupation would be due to interfacial charge transfer from Co to Ni, a Ni$^{2+}$ oxidation state is expected close to the interface. Due to the overlap between the La M$_4$ and the Ni L$_3$ edge, we were not able to resolve the Ni oxidation state using XAS.\\
To block the interaction between the LaNiO$_3$ and LaCoO$_3$ layers, we employed a LaAlO$_3$ spacer layer. The resulting spectrum in \autoref{fig:XAS_Co}a in yellow, shows a predominantly Co$^{3+}$ LS ($\sim$90\%) spectral shape. Hence, we conclude that the LaAlO$_3$ spacer layer successfully blocks electron transfer from Co to Ni. This predominantly LS state is unexpected for LaCoO$_3$ thin films at room temperature \cite{RupanLaCoO3}. A similar effect was observed by Jeong \textit{et al.} \cite{LAOLCOsuperlattice} in LaCoO$_3$-LaAlO$_3$ superlattices, where the authors attribute a Co$^{3+}$ LS state to the reduction of dimensionality. This observation will be revisited in the STEM analysis. \\
A partial Co $d^7$ occupation is expected for the LaTiO$_3$-LaCoO$_3$ interface due to electron transfer from Ti to Co \cite{ZhongHansmann}. By comparison of the experimental spectra with the simulated Co$^{2+}$ spectrum, it is evident that the LaTiO$_3$-LaCoO$_3$ interface contains a high Co$^{2+}$ population. Furthermore, the white line shifts to slightly lower photon energies ($\sim$0.2 eV). The small contributions at $\sim$785 and 799.5 eV are most likely originating from the $M_{4,5}$ edge of a small Ba contamination in the LaTiO$_3$ layer. A linear combination fit of simulated spectra results in an approximate concentration of $\sim$59\% Co$^{2+}$. As shown in \autoref{fig:XAS_Co}, the linear combination deviates slightly in spectral shape from the experimental data due to the relative intensity of the peaks in the simulated Co$^{2+}$ spectrum. These discrepancies are strongly dependent on broadening of the theoretical spectra and are consistent with literature \cite{DeGroot_XASreview,Giashi_LMCO,Schooneveld_CoO}. Qualitatively, the partial $d^7$ occupation is in agreement with the expected charge transfer from Ti to Co across this interface \cite{JaapLTOLCO,ZhongHansmann}. However, a 100\% Co$^{2+}$ population was reported for a 4 u.c. thick LaCoO$_3$ layer between LaTiO$_3$ by Araizi-Kanoutas \textit{et al.}\cite{JaapLTOLCO}, while we find a lower Co$^{2+}$ population ( $\sim$ 59\%). Two potential factors may account for this discrepancy. First of all, the oxygen vacancy content in LaCoO$_3$ is expected to be lower in this work due to the growth pressure, which is two orders of magnitude higher in this work. Second, charge transfer between LaTiO$_3$ and LaNiO$_3$ can result in less charge tranfer at the LaTiO$_3$-LaCoO$_3$ interface. Charge transfer between LaTiO$_3$ and LaNiO$_3$ is expected, as was previously reported by Disa \textit{et al.} \cite{Disa_LTOLNO}, leaving fewer electrons to transfer from LaTiO$_3$ to LaCoO$_3$. When blocking charge transfer between LaTiO$_3$ and LaNiO$_3$ with a LaAlO$_3$ interlayer, we find an approximate population of $\sim$68\% Co$^{2+}$ (Figure \ref{fig:XASCo_LNATCOvsLNTCO}). This confirms that the interaction between LaTiO$_3$ and LaNiO$_3$ influences charge transfer at the LaTiO$_3$-LaCoO$_3$ interface. However, even when blocking the interaction between LaTiO$_3$ and LaNiO$_3$, we find a substantially lower $d^7$ population than found with different PLD conditions in prior work \cite{JaapLTOLCO}.\\
In contrast to the LaTiO$_3$-LaCoO$_3$ interface, where LaTiO$_3$ has a high propensity of overoxidizing \cite{OhtomoLTO,ScheidererLTO}, the LaMnO$_3$-LaCoO$_3$ can be grown at relatively high deposition pressures (0.1 mbar). Due to the high deposition pressure, oxygen vacancies should play a smaller role in the $d$ orbital occupation at this interface. A strong Co$^{2+}$ fingerprint for this interface can be observed in \autoref{fig:XAS_Co}a. Linear combination fitting of simulated spectra results in a slightly lower Co$^{2+}$ concentration for this interface ($\sim$ 51\%) compared to the LaTiO$_3$-LaCoO$_3$ interface  ($\sim$ 59\%). As the O 2\textit{p} band centers are closer for LaMnO$_3$ and LaCoO$_3$ compared to LaTiO$_3$ and LaCoO$_3$, a smaller amount of interfacial charge transfer and therefore a lower $d^7$ population is expected for this interface \cite{ZhongHansmann}. Furthermore, the Co$^{2+}$ fingerprint around 779.2 and 779.8 eV shows a difference in relative intensity compared to the LaTiO$_3$-LaCoO$_3$ interface. This is most likely due to a difference in the Co$^{3+}$ HS contribution, which has a shoulder around 779.5 eV (\autoref{fig:XAS_Co}b.)\\
To investigate whether the additional electrons in the LaCoO$_3$ layer originate from the LaMnO$_3$ layer, we examined the Mn $L_{2,3}$ edges for this sample, as well for a 4 u.c. LaCoO$_3$- 5 u.c. LaMnO$_3$ - 4 u.c. LaCoO$_3$ multilayer and a 10 nm LaMnO$_3$ single layer (Figure \ref{fig:XAS_Mn}). We observe a partial Mn$^{4+}$ occupation by XAS as discussed in supplementary section \ref{sec:SIXAS}, confirming charge transfer from Mn to Co.

\begin{figure*}
\includegraphics[width=0.9\textwidth]{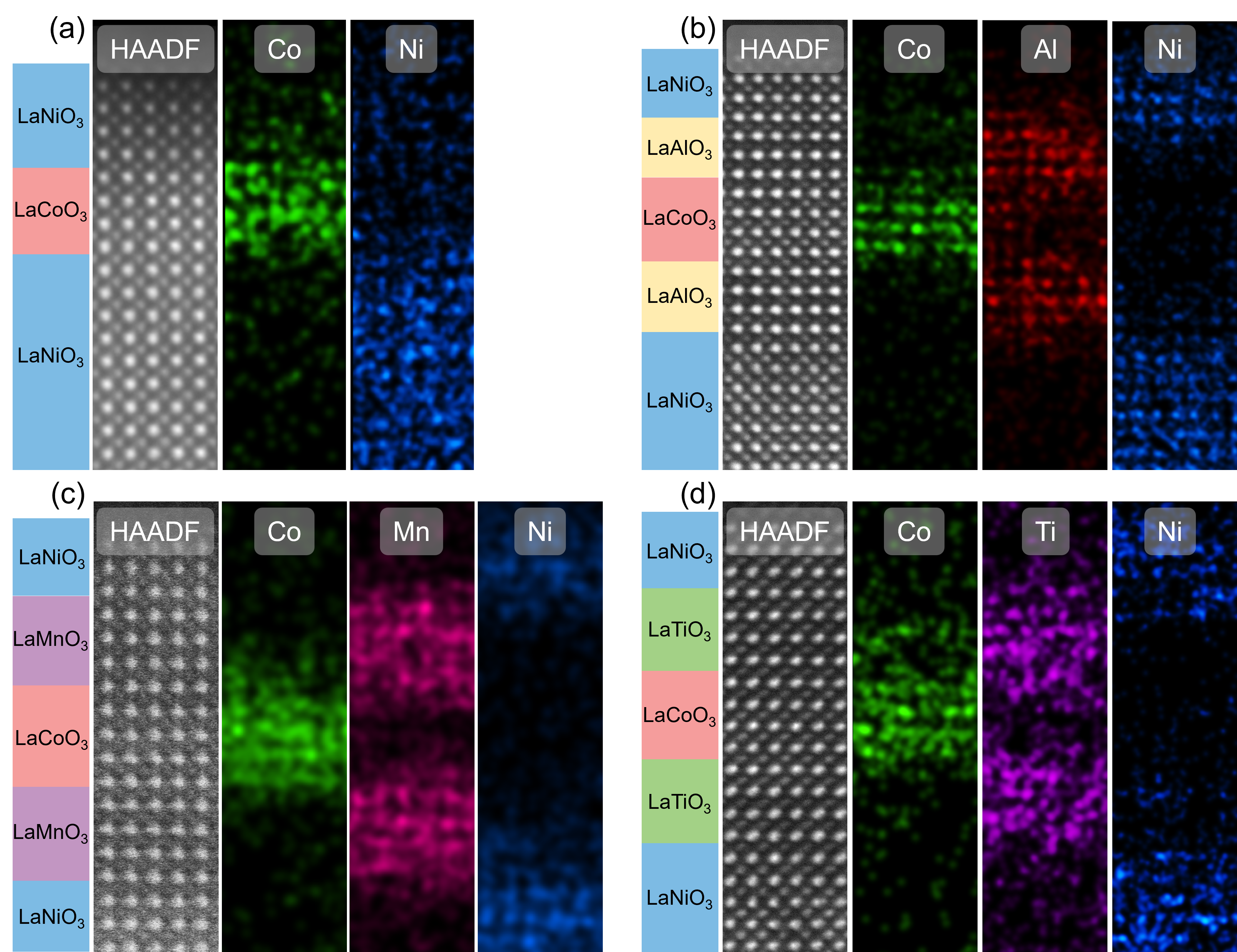}
\caption{\label{fig:TEM_EDX} HAADF STEM images and EDX maps of the multilayers containing the (a) LaNiO$_3$-LaCoO$_3$ interface, (b) LaAlO$_3$-LaCoO$_3$ interface, (c) LaMnO$_3$-LaCoO$_3$ interface, (d) LaTiO$_3$-LaCoO$_3$ interface.}
\end{figure*}

To further examine the origin of the Co valence state close to the interface with different transition metals, we employed STEM to image the interfaces with atomic resolution. EELS (Figure \ref{fig:TEM_EELS}-\ref{fig:EELS_OKedges}) was used to assess the electronic state with spatial resolution. For all samples, the top LaNiO$_3$ layer is likely damaged by X-ray irradiation and lamella preparation and should be merely seen as a protective layer.\\
The EELS maps (Figure \ref{fig:TEM_EELS}) show a rather homogeneous spectral shape for the Co $L$ edge within the signal-to-noise ratio over the thickness of the LaCoO$_3$ layers for all samples. The Co EELS spectra integrated over the full LaCoO$_3$ layer (Figure \ref{fig:EELS_CoLedges}) show roughly the same trend in Co oxidation state as the XAS results (\autoref{fig:XAS_Co}), though with markedly lower signal-to-noise in the EELS data. The spectral weight of the Co $L_3$ edge has higher intensity at lower electron energy loss energies for the samples containing LaTiO$_3$ and LaMnO$_3$ compared to the samples with the LaAlO$_3$ and LaNiO$_3$ interfaces, confirming the substantial Co$^{2+}$ contribution in these samples.\\
The valence states of the transition metals on the other side of the interfaces are also in line with the XAS results. For the sample with the LaNiO$_3$-LaCoO$_3$ interface, the Ni $L_3$ edge overlaps with the La $M_4$ edge, making it difficult to assess the Ni oxidation state. Hence, electron transfer from Co to Ni across the interface cannot be confirmed.\\ 
The EELS Mn $L$ edge of the sample with the LaMnO$_3$-LaCoO$_3$ interface is rather homogeneous over the thickness of the LaMnO$_3$ layers (Figure \ref{fig:TEM_EELS}c). The integrated spectrum over the bottom LaMnO$_3$ layer (Figure \ref{fig:EELS_CoLedges}b) matches the XAS spectrum collected for this sample (Figure \ref{fig:XAS_Mn}), implying a partial Mn$^{4+}$ oxidation state and confirming charge transfer from Mn to Co.\\
For the sample with the LaTiO$_3$-LaCoO$_3$ interface, the Ti EELS signal extends into the LaCoO$_3$ layer (Figure \ref{fig:TEM_EELS}d), pointing towards substantial intermixing for this sample. However, both the Co spectral shape and Ti spectral shape observed with EELS seem rather homogeneous within the signal-to-noise ratio as a function of position. The four peaks in the integrated Ti $L$ edge trace (Figure \ref{fig:EELS_CoLedges}c) indicate a primarily Ti$^{4+}$ oxidation state throughout the structure. Hence, it is likely that the $d^7$ orbital occupation of Co is due to electron transfer from Ti, which is in line with previous work \cite{JaapLTOLCO}.\\
The spatial resolution of EELS also enables us to assess the O $K$ edge per layer (Figure \ref{fig:EELS_OKedges}).  Its spectral shape provides information about the hybridized TM 3$d$ states and their orbital occupation \cite{deGroot_OKedges,Suntivich_OKedges}. However, the O $K$ edges in the LaCoO$_3$ layer show similar spectral shapes as the neighboring layers, as discussed in detail in supplementary section \ref{sec:EELS}.\\

\autoref{fig:TEM_EDX} shows HAADF STEM images for the four interfaces of interest with high-resolution EDX to identify any intermixing at the interfaces. Most interfaces show minimal intermixing within a single unit cell corresponding to the layer roughness during growth. Interestingly, intermixing seems slightly more pronounced at the top of the LaCoO$_3$ layer for all samples, indicating that the LaCoO$_3$ migrates into the layer above. This is in line with previous observations \cite{Wu_LMOLCO_intermixing}.\\
The sample with the LaAlO$_3$-LaCoO$_3$ interfaces (\autoref{fig:TEM_EDX}b) shows a small amount of intermixing of Al into the LaCoO$_3$, barely above the noise level. Substantial mixing of Al in the LaCoO$_3$ structure could result in a change in spin state as LaAl$_{1-x}$Co$_x$O$_3$ with $x = 0.3-0.5$ has been reported to show a decrease in Co$^{3+}$ HS compared to LaCoO$_3$ due to an increase in the crystal field splitting energy \cite{Aswin_LaAlCoO3}. However, cation intermixing alone cannot explain the almost full Co$^{3+}$ LS state observed in the multilayer, as the HS state is expected be populated even for 50\% Al substitution \cite{Aswin_LaAlCoO3} - well beyond the small observable intermixing in \autoref{fig:TEM_EDX}b.\\ 
For the sample with the LaTiO$_3$-LaCoO$_3$ interface, EDX shows a limited amount of intermixing of Ti into the LaCoO$_3$ layer, which is in line with the EELS results, indicating that mixing of the cations may substantially contribute to the charge transfer in this sample.\\ 
The sample with the LaMnO$_3$-LaCoO$_3$ interface (\autoref{fig:TEM_EDX}c) shows a small amount of cation intermixing between LaCoO$_3$ and the top LaMnO$_3$ layer. Electron transfer from Co to Mn is previously reported in LaMn$_{1-x}$Co$_{x}$O$_3$ \cite{Giashi_LMCO}. Hence, the observed Co$^{2+}$ contribution in XAS can originate (partially) from intermixing at this interface \cite{Wu_LMOLCO_intermixing}.\\
As discussed above, EELS indicates that the cation $L$ edges are rather homogeneous in spectral shape over the layer thicknesses and similar for the top and bottom interfaces, while intermixing is stronger close to the interfaces, implying that the charge transfer extends further than intermixing. Thus, the spatial extent of the electronic effect exceeds that of the chemical intermixing, emphasizing that the observed orbital occupancies and spin states result from an intrinsic interfacial phenomenon governed by electronic—not compositional—length scales, and pointing to electronic driving forces.

\begin{figure*}
\includegraphics[width=\textwidth]{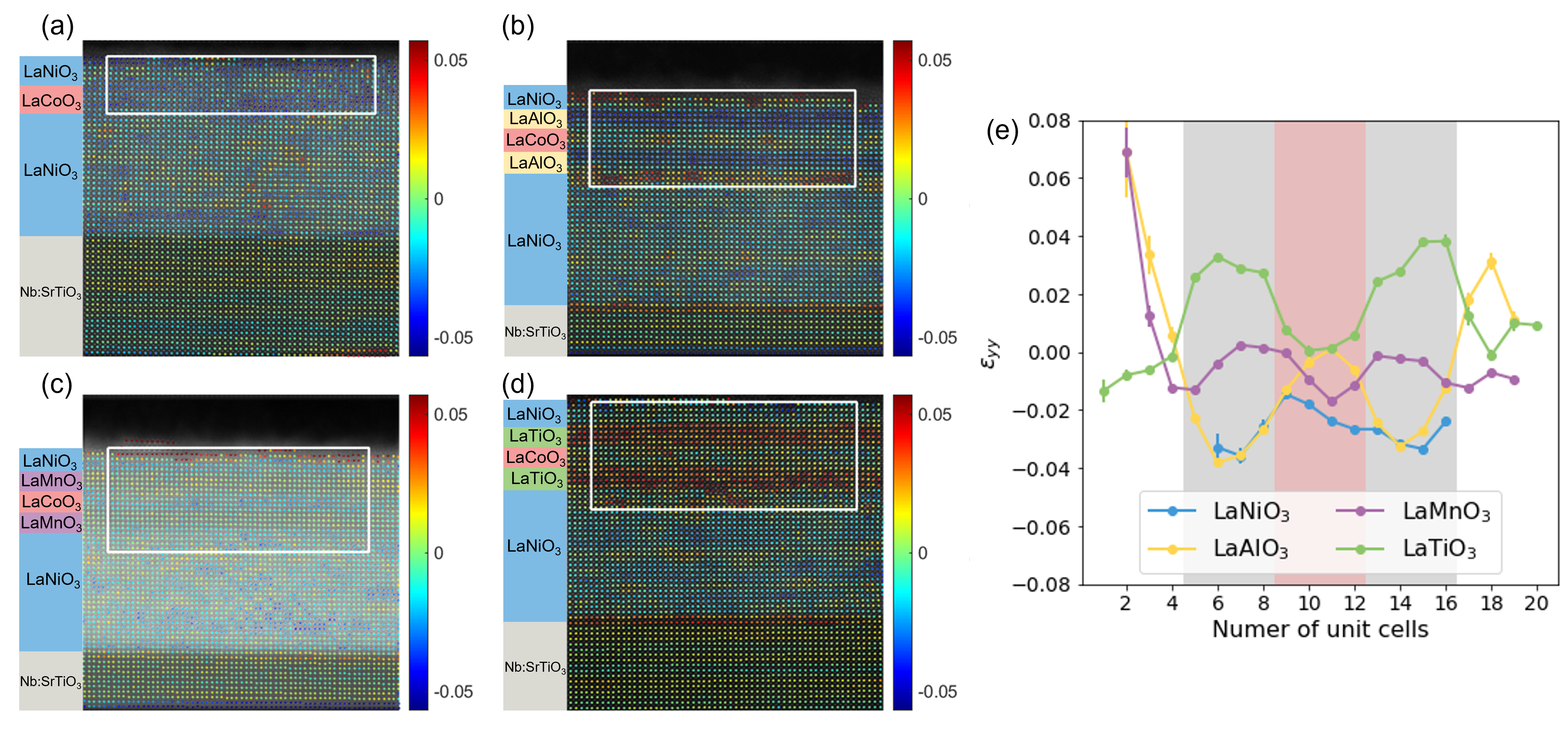}
\caption{\label{fig:TEM_strain} Out-of-plane strain mapping of the multilayers containing the (a) LaNiO$_3$-LaCoO$_3$ interface, (b) LaAlO$_3$-LaCoO$_3$ interface, (c) LaMnO$_3$-LaCoO$_3$ interface, (d) LaTiO$_3$-LaCoO$_3$ interface. White outlined regions are used to calculate the average out of plane strain per layer. (e) Average out of plane strain as a function of distance from the interface. Averaging is performed horizontally across the white area outlined in (a)-(d). The LaCoO$_3$ layer is indicated by the red shaded region, while the neighboring layers are indicated by the grey shaded region. The white region corresponds to the LaNiO$_3$. Error bars indicate standard errors. The legend labels indicate the neighboring layer to the LaCoO$_3$ layer.}
\end{figure*}

While differences in Co valence state can be explained by interfacial charge transfer, the origin of variations in spin remain unclear. Strain analysis reveals more insight into the possible origin of the variations in Co spin states for the different interfaces. All thin films are commensurately tensile strained to the SrTiO$_3$ substrate, resulting in the same in-plane lattice parameter (\autoref{fig:XRDresults}b-e). Typically, tensile strain reduces the out-of-plane lattice parameter resulting in negative out-of-plane strain. Strain analysis by STEM (\autoref{fig:TEM_strain}) shows that the out-of-plane strain deviates for the different interfaces. The largest negative out-of-plane strain is observed for the sample with the LaNiO$_3$-LaCoO$_3$ interface, which contains a high amount of Co$^{3+}$ HS concentration. Upon creation of a LaAlO$_3$-LaCoO$_3$ interface, the out-of-plane strain in the LaCoO$_3$ layer becomes substantially smaller and a primarily Co$^{3+}$ LS population is observed. When comparing the samples with the LaMnO$_3$-LaCoO$_3$ and LaTiO$_3$-LaCoO$_3$ interfaces, both having a large Co$^{2+}$ HS population, we observe a similar effect for the Co$^{3+}$ spin state. The Co$^{3+}$ population has relatively more HS for the LaMnO$_3$-LaCoO$_3$ interface, which exhibits more negative out-of-plane strain in the LaCoO$_3$ layer. Thus, we observe that LaCoO$_3$ layers with negative out-of-plane strain have more Co$^{3+}$ HS, and samples with little out-of-plane strain have more Co$^{3+}$ LS.\\
The Co$^{3+}$ spin state is known to be closely related to strain \cite{Sterbinsky_LCOexafs,RupanLaCoO3,Mehta_LCOmagnetismvacancies,Gupta_LCOstraindirvenmagnetism}. Tetragonal distortions due to in-plane tensile strain are known to lift the degeneracy of the 3$d$ orbitals, thereby decreasing the crystal field splitting energy and favoring Co$^{3+}$ HS population. The observed out-of-plane strain for the samples showing more Co$^{3+}$ HS (LaNiO$_3$-LaCoO$_3$ and LaMnO$_3$-LaCoO$_3$ interfaces) points towards such a tetragonal distortion that decreases the crystal field splitting energy and results in a favored HS population. For the samples that show more Co$^{3+}$ LS (LaAlO$_3$-LaCoO$_3$ and LaTiO$_3$-LaCoO$_3$ interfaces), contrary to the usual epitaxial strain response, we find minimal negative out-of-plane strain. This leads to a higher crystal field splitting energy compared to the HS samples, which favors occupation of the t$_{2g}$ orbitals over the e$_g$ orbitals and thereby a Co$^{3+}$ LS state. \\
Thus, strain analysis combined with spectroscopy of a set of La$B$O$_3$-LaCoO$_3$ interfaces points towards the importance of structural distortions and strain relaxation mechanisms in the \textit{d} orbital occupation of Co in LaCoO$_3$. Strain relaxation provides an alternative argument to the dimensionality argument provided by Jeong \textit{et al.} \cite{LAOLCOsuperlattice}. In their work, the authors observed a predominantly Co$^{3+}$ LS state in LaAlO$_3$-LaCoO$_3$ superlattices as the reduced dimensionality of the LaCoO$_3$ layers between insulating LaAlO$_3$ increased the crystal field splitting energy compared to the bulk.\\ 
Unfortunately, the exact nature of the LS state in our multilayers remains unknown without further investigation of the strain relaxation mechanism. Nevertheless, these findings demonstrate that the electronic ground state (oxidation state and spin state) in correlated oxide interfaces cannot be understood from charge transfer or dimensionality alone; rather, it emerges from the coupled evolution of strain, symmetry distortion, and interfacial structural accommodation.

\section{\label{sec:conclusions}Conclusions}
We demonstrated control over the $d$ orbital occupation in LaCoO$_3$ by the use of interfacial engineering. Using CTM calculations to interpret the Co $L$ edges, we observe a partial $d^7$ occupation at the interface with LaTiO$_3$. A partial $d^7$ orbital occupation is also formed at the interface with LaMnO$_3$, which does not show the same tendency to overoxidize as LaTiO$_3$. Hereby, we confirm that oxygen nonstoichiometry is not the main driving force of charge transfer at these isopolar interfaces. Furthermore, we found a partial Co $d^5$ occupation at the interface with LaNiO$_3$, and blocking of this charge transfer with a LaAlO$_3$ layer. Aside from a change in valence state due to interfacial charge transfer, we observe a change in spin state for the different interfaces. Strain analysis of high-resolution STEM images points towards a correlation between out-of-plane strain and the Co$^{3+}$ spin state,  identifying that structural distortions  - in addition to reduced dimensionality and interfacial charge transfer - dictate the spin-state of LaCoO$_3$, underscoring the multifaceted nature of electronic reconstruction at oxide interfaces. Control over the $d$ orbital occupancy through interfacial engineering offers a powerful and underappreciated design strategy for oxide electronics and (electro)catalysis, where the outer-electron orbital occupancy and spin states dictate key material properties and functionality.
\begin{acknowledgments}
This work is funded by the Netherlands Ministry of Economic Affairs' Top Consortia for Knowledge and Innovation (TKIs) Allowance (CHEMIE.PGT.2021.007). This work was carried out with the support of Diamond Light Source, instruments B07-B and B07-C (proposal SI33107 and cm31118). The authors thank Pilar Ferrer for her support during the XAS experiments. The authors thank Guido Mul for fruitful discussions.
\end{acknowledgments}

\section*{Author contributions}
E.K. synthesized the samples and performed RHEED, XRD and AFM measurements. E.K., E.M., I.B. and M.S. performed XAS measurements. E.K. performed CTM simulations and analysed the XAS data. F.G. supervised the simulations and XAS data analysis. N.G. performed STEM measurements. N.G., Z.Z. and A.A. analysed EDX and EELS data. J.V. supervised the EELS/EDX analysis. A.D.B. performed STEM strain analysis. S.V.A. supervised the STEM strain analysis. C.B., B.M. and G.K. conceptualized the project and supervised the research. E.K. wrote the manuscript with contributions from all authors. All authors discussed the results and contributed to the final version.

\section*{References}
\bibliography{references.bib}

\newpage
\begin{widetext}

\appendix

\renewcommand{\appendixname}{}%

\renewcommand{\thesection}{S\arabic{section}}
\renewcommand{\thesubsection}{S\arabic{section}.\arabic{subsection}}
\setcounter{section}{0}
\setcounter{figure}{0}
\setcounter{table}{0}

\section*{Supplemental material}

\makeatletter
\renewcommand \thesection{S\@arabic\c@section}
\renewcommand\thetable{S\@arabic\c@table}
\renewcommand \thefigure{S\@arabic\c@figure}
\makeatother

\renewcommand{\thesection}{S\arabic{section}}
\renewcommand{\thesubsection}{S\arabic{section}.\arabic{subsection}}
\setcounter{section}{0}


\section{Pulsed laser deposition parameters}

\begin{table}[h!]
\caption{\label{tab:deppar} PLD growth parameters for the different layers. The substrate-target distance was 50 mm and the spot size of the ablated area was 1.8 mm$^2$ for all layers.}
\begin{ruledtabular}
\begin{tabular}{ccccc}
 & Fluence (J/cm$^2$) & Temperature ($\degree$C) & $pO_2$ (mbar) & Repetition rate (Hz)\\
\hline
LaNiO$_3$ & 1.8 & 450 & 0.04 & 2\\
LaCoO$_3$ & 1.8 & 670 & 0.1 & 2\\
LaMnO$_3$ & 1.8 & 670 & 0.1 & 1\\
LaTiO$_3$ & 1.8 & 670 & $2\cdot 10^{-3}$ & 1\\
LaAlO$_3$ & 1.3 & 670 & $2\cdot 10^{-3}$ & 1\\

\end{tabular}
\end{ruledtabular}
\end{table}

\section{\label{sec:CTMcalc} Charge transfer multiplet calculations parameters}
The semi-empirical simulation parameters for Co$^{2+}$ HS, Co$^{3+}$ LS, Co$^{3+}$ HS and Co$^{4+}$ HS are similar to the parameters reported in literature \cite{TomiyasuLCO,RupanLaCoO3,Giashi_LMCO}. The crystal field splitting, $10D_q$, for Co$^{3+}$ LS and HS were optimized to match experimental spectra of EuCoO$_3$ and Sr$_2$CoO$_3$Cl \cite{EuCoO3}. The $10D_q$ value of Co$^{2+}$ HS was optimized match the distance between first three peaks of the Co$^{2+}$ feature in the L$_3$ edges presented in this work. The parameters for Co$^{4+}$ were optimized to match the experimental spectrum of SrCoO$_3$ \cite{Li_LCOSCOref}. As the calculations do not give absolute energy positions, the spectra are shifted in energy using the experimental spectra listed above.

\begin{table}[h!]
\caption{\label{tab:XASsimuCopar} Parameters used for charge transfer multiplet calculations of Co $L_{2,3}$ edges. The total crystal field is a combination of the ionic value $10D_q$ (ionic) and a covalent contribution that is approximately 0.3 for all cases. The $10D_q$ is reduced by $\sim$ 20\% in the fnial state with respect to the ground state.} 
\begin{ruledtabular}
\begin{tabular}{ccccccccccc}
 & \multicolumn{2}{c}{Crystal field}&\multicolumn{3}{c}{Slater reduction} & \multicolumn{5}{c}{Charge transfer}\\
 \hline
 & Symmetry & $10D_q$ (ionic) & F$_{dd}$ & F$_{pd}$ & G$_{pd}$ & $\Delta$ & U$_{dd}$ & $U_{pd}$ & T$_{e_g}$ & T$_{t_2g}$\\
 \hline
Co$^{2+}$ HS & O$_h$ & 0.85 & 0.9 & 0.9 & 0.9 & 3.0 & 6.5 & 7.5 & 2 & 1\\
Co$^{3+}$ LS & O$_h$ & 1.3 & 0.8 & 0.8 & 0.8 & 1.5 & 6.5 & 7.5 & 2 & 1\\
Co$^{3+}$ HS & O$_h$ & 0.8 & 0.8 & 0.8 & 0.8 & 1.5 & 6.5 & 7.5 & 2 & 1\\
Co$^{4+}$ HS & O$_h$ & 1.6 & 0.8 & 0.8 & 0.8 & -1.5 & 6.5 & 7.5 & 2 & 1\\
\end{tabular}
\end{ruledtabular}
\end{table}

For Mn$^{2+}$, parameters similar to reference \citenum{degroot_Mn2ref} are chosen. The symmetry of Mn$^{3+}$ is chosen to be D$_{4h}$ rather than O$_h$ as Jahn-Teller distortions tend to result in a small tetragonal distortion in the crystal field \cite{Herrera2023,Chaluvadi_LSMO}. To model the high covalency of Mn$^{4+}$ , the Slater integrals were reduced further from their atomic values \cite{Radtke_MnSlaterreductie}. The simulation parameters chosen for Mn$^{3+}$ and Mn$^{4+}$ are similar to the parameters chosen in references \citenum{Chaluvadi_LSMO} and \citenum{Giashi_LMCO}. The relative energy positions of each valence state are shifted to match reference spectra of MnO, Mn$_2$O$_3$ and MnO$_2$ \cite{Gilbert_MnXASrefs}.

\begin{table}[h!]
\caption{\label{tab:XASsimuMnpar} Parameters used for charge transfer multiplet calculations of Mn $L_{2,3}$ edges. The total crystal field is a combination of the ionic value $10D_q$ (ionic) and a covalent contribution that is approximately 0.3 for all cases. The $10D_q$ is reduced by $\sim$ 20\% in the fnial state with respect to the ground state.}
\begin{ruledtabular}
\begin{tabular}{ccccccccccc}
 & \multicolumn{2}{c}{Crystal field}&\multicolumn{3}{c}{Slater reduction} & \multicolumn{5}{c}{Charge transfer}\\
 \hline
 & Symmetry & $10D_q$ (ionic) & F$_{dd}$ & F$_{pd}$ & G$_{pd}$ & $\Delta$ & U$_{dd}$ & $U_{pd}$ & T$_{e_g}$ & T$_{t_2g}$\\
 \hline
Mn$^{2+}$ & O$_h$ & 0.8 & 0.9 & 0.9 & 0.9 & - & - & - & - & - \\
Mn$^{3+}$ & D$_{4h}$ \footnote{D$_t$=-0.02, D$_s$ = 0.1} & 1.55 & 0.8 & 0.8 & 0.8 & 4 & 6.5 & 7.5 & 2 & 1 \\
Mn$^{4+}$ & O$_h$ & 2.1 & 0.4 & 0.5 & 0.5 & 1 & 6.5 & 7.5 & 2 & 1 \\
\end{tabular}
\end{ruledtabular}
\end{table}

\clearpage

\section{Thin film characterization}

\begin{figure*}[h!]
\includegraphics[width=\textwidth]{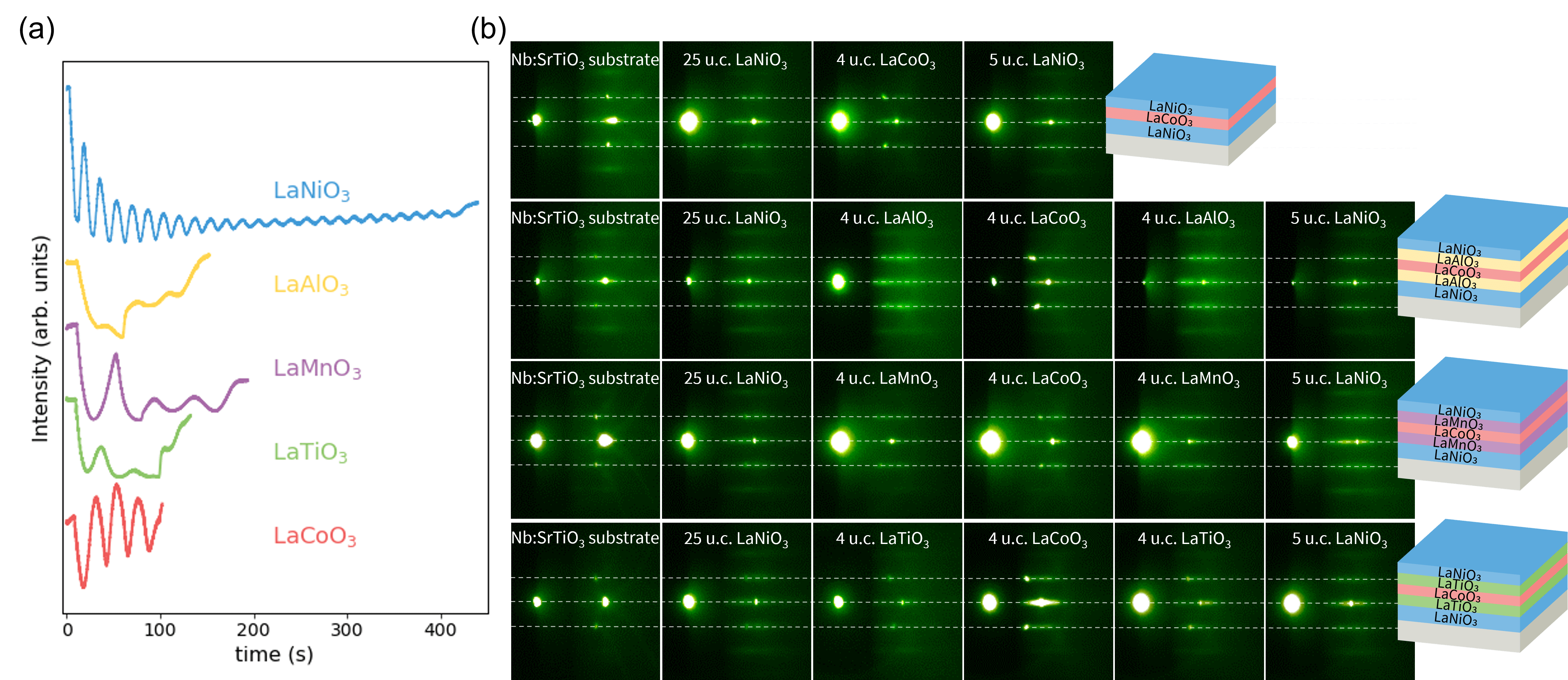}
\caption{\label{fig:rheedSI} a) Integrated intensity of the specular spot as a function of time during the deposition of each layer. Sharp increases are due to a manual increase of the filament current. b) RHEED patterns after the deposition of each layer.}
\end{figure*}

\begin{figure*}[h!]
\includegraphics[width=0.7\textwidth]{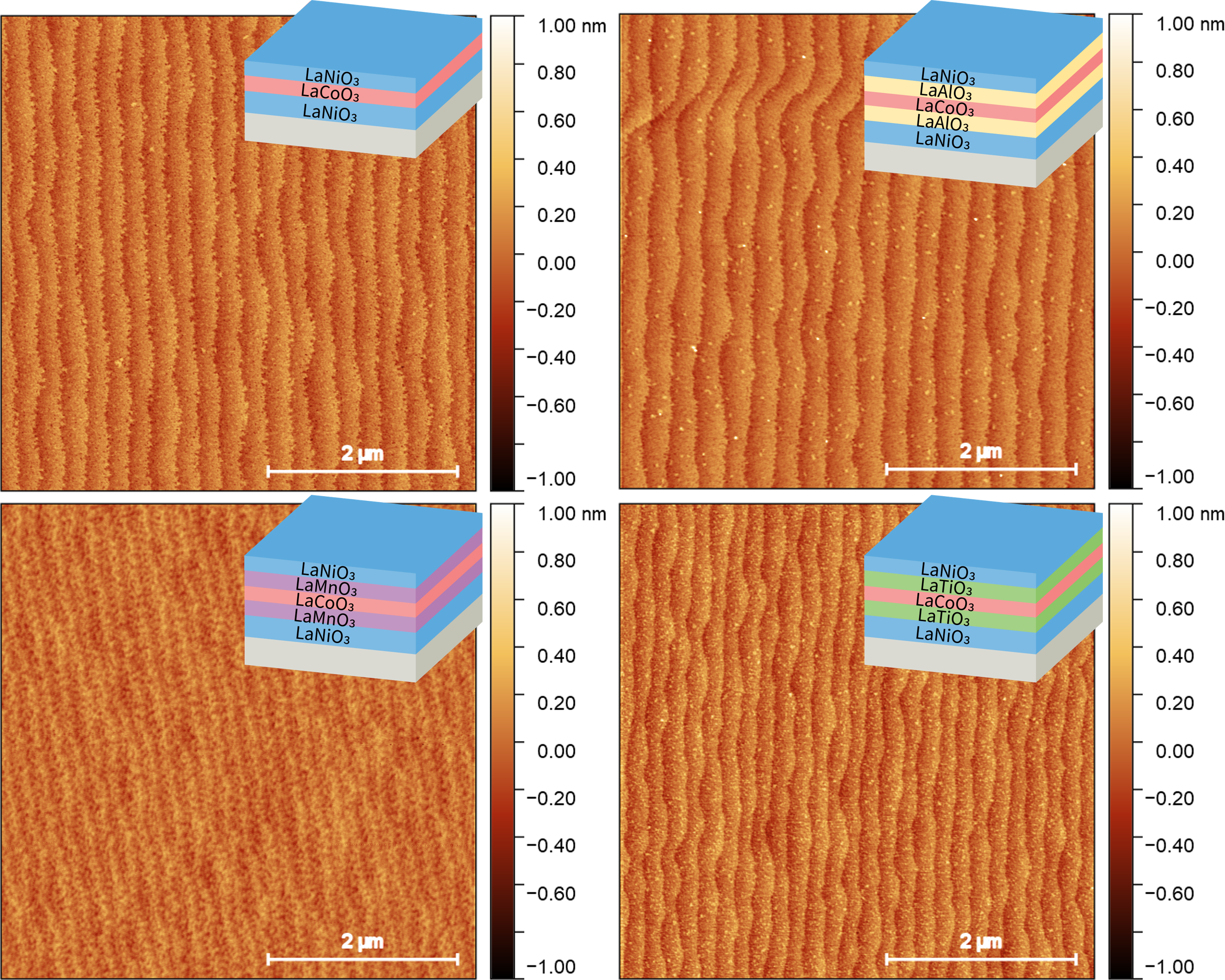}
\caption{\label{fig:AFMresults} AFM images of the LaCoO$_3$ multilayers with the (a) LaNiO$_3$-LaCoO$_3$ interface, (b) LaAlO$_3$-LaCoO$_3$ interface, (c) LaMnO$_3$-LaCoO$_3$ interface, (d) LaTiO$_3$-LaCoO$_3$ interface.}
\end{figure*}

\clearpage
\newpage
\section{\label{sec:SIXAS}Additional XAS data and fitting}

\begin{figure*}[h!]
\includegraphics[width=\textwidth]{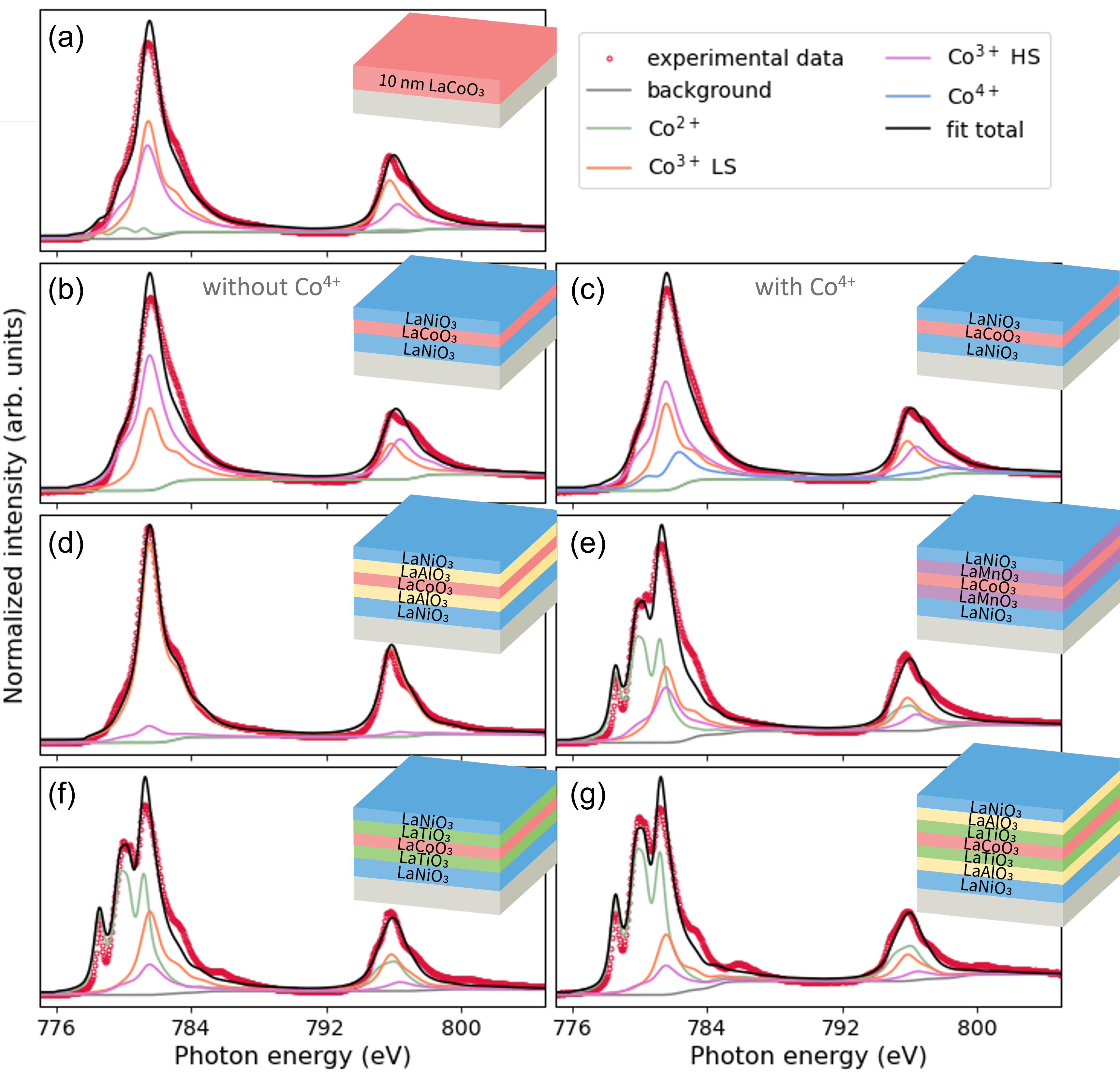}
\caption{\label{fig:XASCofits} Linear combination fits of calculated spectra for different Co valence states.}
\end{figure*}

\begin{figure*}[h!]
\includegraphics[width=0.6\textwidth]{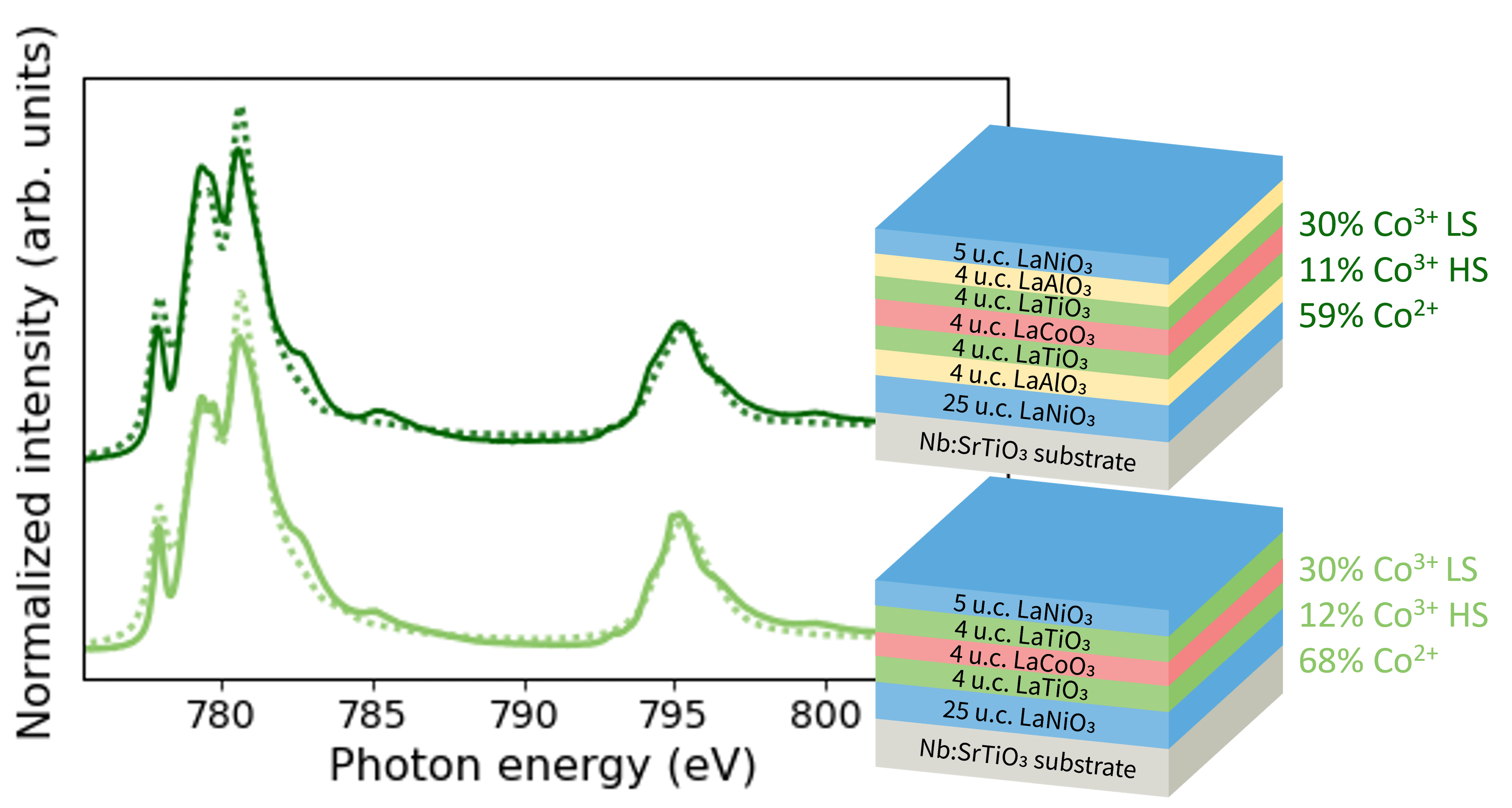}
\caption{\label{fig:XASCo_LNATCOvsLNTCO} Co $L_{2,3}$ edges with and without a LaAlO$_3$ spacer between LaTiO$_3$ and LaNiO$_3$ (solid lines). The dotted lines represent a linear combination fit of simulated spectra using CTM calculations. Numbers on the right side indicate relative contributions of simulated spectra to fit.}
\end{figure*}

\newpage
\subsection*{Mn $L$ edges of the LaCoO$_3$-LaMnO$_3$ interface}
XAS spectra of the sample with the LaCoO$_3$-LaMnO$_3$ interface and a single LaMnO$_3$ layer are shown in figure \ref{fig:XAS_Mn}. Furthermore, a multilayer with a single LaMnO$_3$ layer in between two LaCoO$_3$ layers is included. Both samples with a LaCoO$_3$-LaMnO$_3$ interfaces show a shift in photon energy, implying a higher oxidation state compared to a single LaMnO$_3$ layer. Comparison of the experimental spectra with calculated spectra of Mn$^{2+}$, Mn$^{3+}$ and Mn$^{4+}$ are shown in figure \ref{fig:XAS_Mn}b to help identify the differences between experimental spectra. As Mn $L$ edges are highly affected by Jahn-Teller distortions and covalency effects \cite{Herrera2023}, we refrain from fitting the calculations to the experimental spectra. However, the shift in photon energy and the change in spectral shape indicate a higher Mn$^{4+}$ contribution for the multilayers compared to the single layer. Interestingly, a layer of LaMnO$_3$ between two LaCoO$_3$ layers results in a higher Mn$^{4+}$ contribution than the sample with  LaCoO$_3$ between two LaMnO$_3$ layers, pointing towards more charge transfer between Mn and Co compared to Ni and Co. 

\begin{figure*}[h!]
\includegraphics[width=0.5\textwidth]{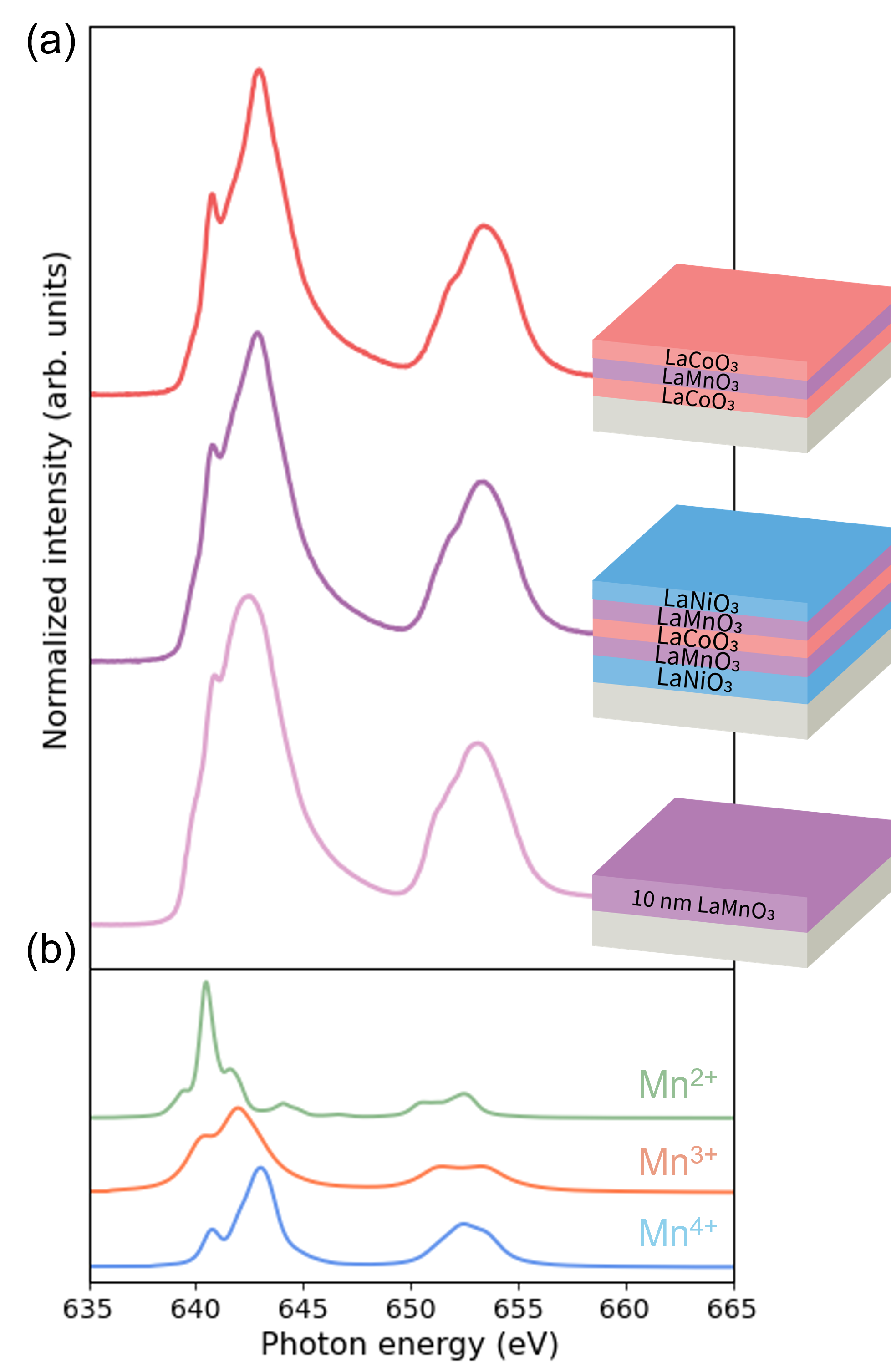}
\caption{\label{fig:XAS_Mn} (a) Mn L$_{2,3}$ edges of multilayers with LaCoO$_3$-LaMnO$_3$ interfaces as well as a single LaMnO$_3$ layer. (b) CTM calculations of Mn$^{2+}$, Mn$^{3+}$ and Mn$^{4+}$ valence states.}
\end{figure*}

\clearpage
\newpage
\section{\label{sec:EELS}EELS data}

\begin{figure*}[h!]
\includegraphics[width=0.9\textwidth]{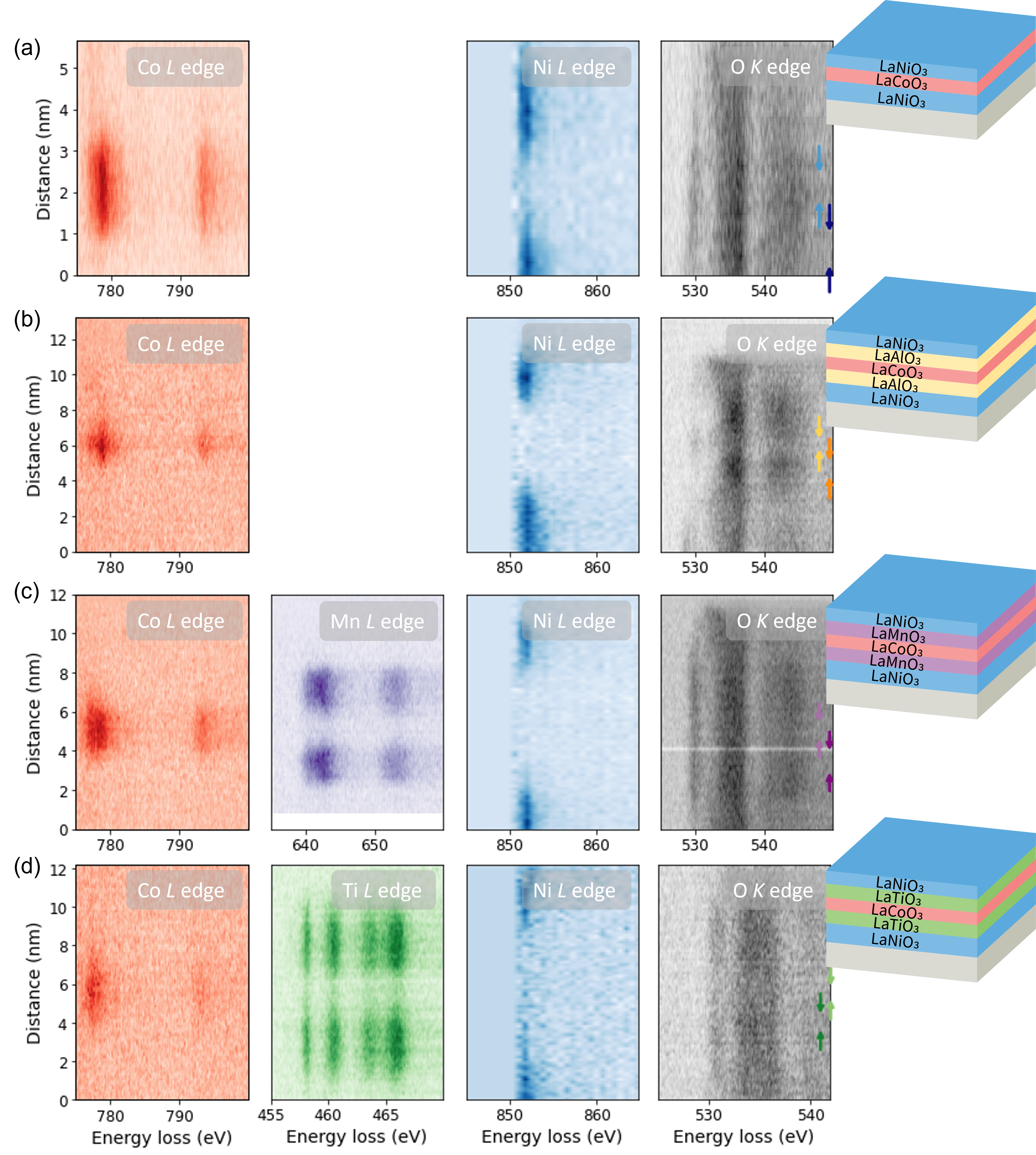}
\caption{\label{fig:TEM_EELS} EELS of the TM $L$ edges and O K edges in the multilayers containing the (a) LaNiO$_3$-LaCoO$_3$ interface, (b) LaAlO$_3$-LaCoO$_3$ interface, (c) LaMnO$_3$-LaCoO$_3$ interface, (d) LaTiO$_3$-LaCoO$_3$ interface. The Mn $L$ edge in (c) is taken at a different but equivalent position. The data is shifted over the spatial axis such that the Co and Ni signals match the between the two positions. The arrows indicate integration regions for the spectra in figure \ref{fig:EELS_CoLedges} and \ref{fig:EELS_OKedges}.}
\end{figure*}

\begin{figure*}[h!]
\includegraphics[width=\textwidth]{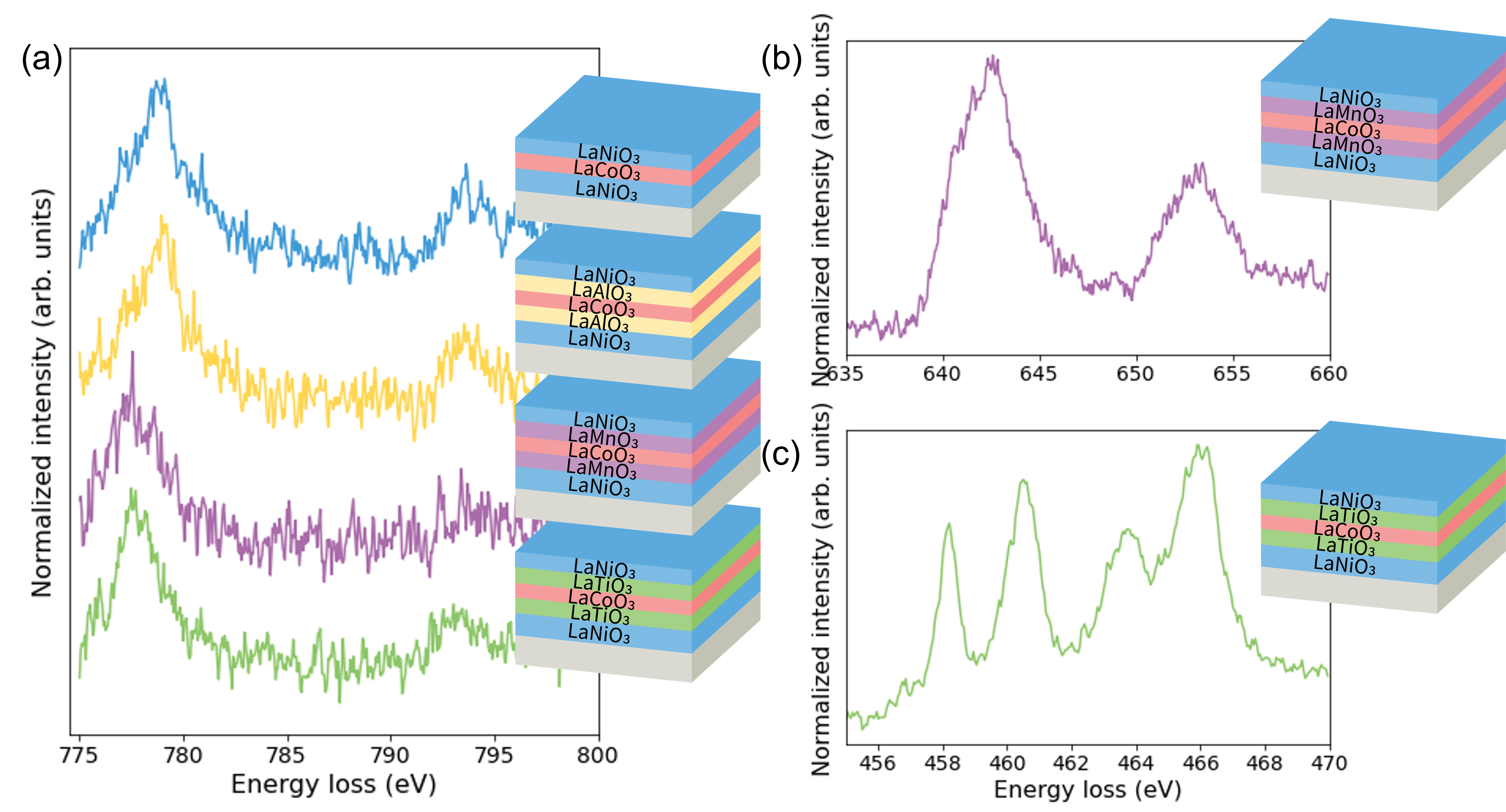}
\caption{\label{fig:EELS_CoLedges}(a) EELS traces of the Co $L$ edges for the different multilayers extracted from the EELS maps by integrating for the full LaCoO$_3$ layer. (b) Mn $L$ edge for the LaMnO$_3$-LaCoO$_3$ multilayer integrated for the bottom LaMnO$_3$ layer. (c) Ti $L$ edge LaTiO$_3$-LaCoO$_3$ multilayer integrated for the bottom LaTiO$_3$ layer.}
\end{figure*}

\subsection*{O K edges}

Figure \ref{fig:EELS_OKedges} shows the O K edges extracted at the center of the LaCoO$_3$ layer (solid lines) and the layer below (dotted lines) for each of the multilayers. The correponding integrated regions in the EELS maps are indicated by the colored arrows in figure \ref{fig:TEM_EELS}. We find that the pre-edge at $\sim$ 530 eV is very low in intensity for the sample with the LaAlO$_3$-LaCoO$_3$ interface. The pre-edge is related to electron excitations from the O 1$s$ state to the hybridized O 2$p$-TM 3$d$ state and therefore provides information about the hybridized TM 3$d$ states\cite{deGroot_OKedges,Suntivich_OKedges}. Due to the empty $d$ orbitals of LaAlO$_3$, no significant hybridization and therefore no pre-edge feature is expected in these layers. However, this electronic state seems to extend into the LaCoO$_3$ layer. Furthermore, the pre-edge position is very similar inside the LaCoO$_3$ layer compared to the layer below for the samples with the LaMnO$_3$-LaCoO$_3$ and LaTiO$_3$-LaCoO$_3$ as well. Taken together, these observations suggest that the signal of the O K edges is more delocalized than the signal from the cations and exceeds the energy loss delocalization of approximately 4 \AA. A similar effect was previously observed for LaNiO$_3$-LaAlO$_3$ superlattices \cite{Nicolas_LNOLAO}. We hypothesize that the delocalization of the electronic states of oxygen is more prominent compared to the cations due to the hybridized nature of the O 2$p$ states. Therefore, we use TM edges for charge transfer evaluation, which cannot be resolved with the more delocalized O K edges.\\

\begin{figure*}[h]
\includegraphics[width=0.5\textwidth]{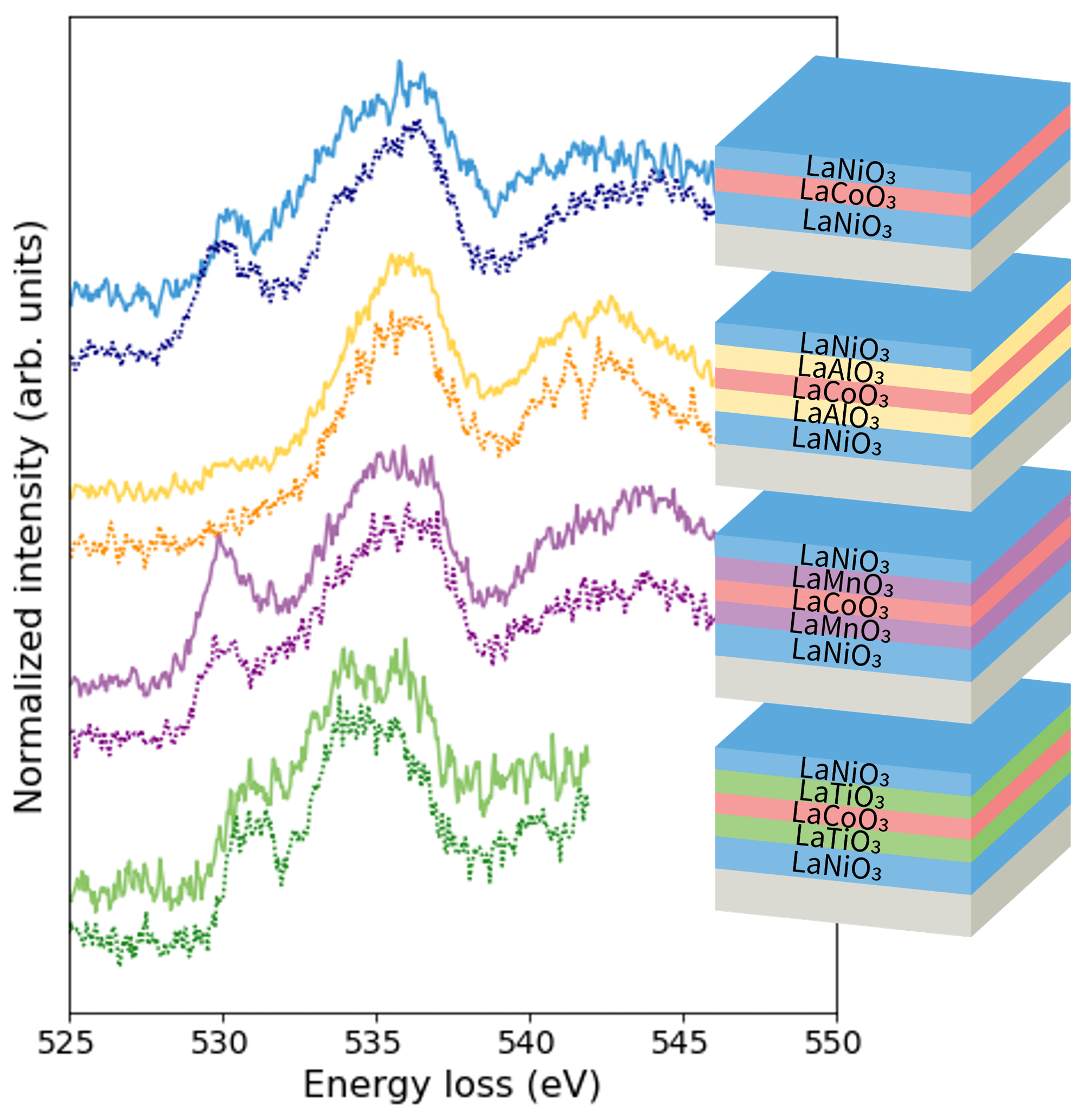}
\caption{\label{fig:EELS_OKedges} EELS traces of the O K edges for the different multilayers extracted from the EELS maps in the center of the LaCoO$_3$ layer (solid lines) and layer below (dotted lines) as indicated by the colored arrows in figure \ref{fig:TEM_EELS}.}
\end{figure*}

\end{widetext}


\end{document}